\documentclass{article}
\usepackage{spconf,amsmath,graphicx}

\usepackage{amssymb}
\usepackage{booktabs}
\usepackage{color}
\usepackage{enumitem}
\usepackage{multirow}
\usepackage[skip=2pt]{caption}
\usepackage[table]{xcolor}  

\newcommand{\wv}{wav2vec~2.0 }
\newcommand{\torgo}{the TORGO database}
\newcommand{\uaspeech}{UASpeech }

\definecolor{myb}{RGB}{96, 125, 163}
\definecolor{myy}{RGB}{212, 189, 47}
\colorlet{myo}{orange}
\colorlet{myr}{red}
\usepackage[breaklinks=true,colorlinks,bookmarks=false]{hyperref}

\title{Speech Intelligibility Classifiers from 550k Disordered Speech Samples}
\name{\vspace{-0.4em}\begin{tabular}{c}Subhashini Venugopalan$^{1}$, Jimmy Tobin$^1$, Samuel J. Yang$^{1}$, Katie Seaver$^{1,3}$, Richard J.N. Cave$^1$, \\ Pan-Pan Jiang$^1$, Neil Zeghidour$^{2}$, Rus Heywood$^1$, Jordan Green$^{1,3}$, Michael P. Brenner$^{1,4}$ \end{tabular}}

\address{
  $^1$Google Research, USA; 
  $^2$Google Research, France; \\
  $^3$MGH Institute of Health Professions, USA;
  $^4$Harvard University, USA}
\begin{document}
\maketitle
\begin{abstract}
We developed dysarthric speech intelligibility classifiers on 551,176 disordered speech samples  contributed by a diverse set of 468 speakers, with a range of self-reported speaking disorders and rated for their overall intelligibility %
on a five-point scale. 
We trained three models following different deep learning approaches and evaluated them on ${\sim}$94K utterances from 100 speakers.
We further found the models to generalize well (without further training) on
\torgo\cite{rudzicz2012torgo} (100\% accuracy),  UASpeech\cite{kim2008dysarthric} (0.93 correlation), ALS-TDI PMP\cite{vieira2022machine} (0.81 AUC) datasets %
as well as on a dataset of 
realistic unprompted speech we gathered (106 dysarthric and 76 control speakers,
  ${\sim}$2300 samples). 
\end{abstract}

\section{Introduction}
\vspace{-0.1cm}
Atypical speech can manifest from a variety of conditions. Neurological diseases such as Amyotrophic Lateral Sclerosis (ALS), Parkinson's Disease (PD), and Cerebral Palsy (CP), are amongst the most prevalent causes of dysarthria and speech disability. %
Automatic assessments of speech intelligibility can help predict how well voice-based assistive technologies might aid a person with speech disorders~\cite{rudzicz2012torgo}. They can be used to detect such speech e.g.~in YouTube, to allow better transcriptions from  specialized Automatic Speech Recognition (ASR) systems~\cite{euphonia1Mdata}, or used by researchers as an objective measure to monitor decline in speech e.g., in ALS~\cite{vieira2022machine}. 
Such classifiers can also help identify variable manifestations of impaired speech, to enable automatic collection of such data at scale to teach and improve ASR systems. %

Classification of speech disorders and in particular classifying dysarthric speech and speech intelligibility have been fairly well studied for different applications~\cite{huang2021review,kim2015automatic}. Many works have developed machine learning models based on handcrafted acoustic features~\cite{hansen1998nonlinear,baghai2012automatic,khan2014classification,liu2018gmm}. Among deep learning methods, convolutional neural networks (CNNs) are quite popular~\cite{liu2018gmm,janbakhshi2020automatic,TRILL}, as are recurrent neural networks (RNNs), specifically Long-Short-Term-Memory (LSTM)~\cite{lstm}   models~\cite{mayle2019diagnosing,kim2018dysarthric,millet2019learning} have also been used to classify dysarthric speech.
Some recent works have explored transformer~\cite{vaswani2017attention,gulati2020conformer} based models for non-speech classification tasks~\cite{zhang2021bigssl,shor2021universal}. However, most prior works developed models on much smaller datasets of disordered speech, with fewer utterances and speakers, and focused on a limited set of phrases or speech disorder etiologies.
We use a large dataset of 756,147 utterances contributed by 677
speakers with a range of self-reported speech disorders as part of \textit{Project Euphonia}~\cite{euphonia1Mdata}. The speech samples are rated for their overall intelligibility on a five-point Likert scale by speech-language pathologists (SLPs). We build classification models based on different deep learning architectures including convolutional networks with learnable audio frontends~\cite{zeghidour2021leaf}, representations from an LSTM-based ASR encoder model~\cite{venugopalan2021comparing}, and representations from the self-supervised \wv CNN and transformer architecture backbone~\cite{baevski2020wav2vec}. The models are trained on over 550K samples
to predict either a binary (typical, atypical speech) label, or the five class labels. The models achieved an accuracy of over 86\% when  evaluated on a test set of nearly 94K utterances from 100 speakers.
To assess the flexibility and generalizability of the models, we also evaluated (inference only) %
on (1) \torgo~\cite{rudzicz2012torgo} consisting of %
14 speakers; %
(2) the \uaspeech dataset~\cite{kim2008dysarthric} with 28 speakers; %
 and (3) the ALS-TDI PMP dataset~\cite{vieira2022machine} with 90 speakers; and (4) on unconstrained realistic speech gathered from videos of 76 controls and 106 dysarthric speakers covering 5 etiologies. Our models showed performance competitive with the state-of-the-art (SOTA) on all datasets. We found it to perform well on speakers with ALS, PD, CP, and Ataxia. We describe our models and evaluation, and share our findings here. 

\section{Disordered speech classification}

Our work focuses on classifying {\sl intelligibility} of dysarthric speech. 
Intelligibility measures how well speech is understood by a human listener \cite{stipancic2018minimally}. 
In our dataset (Sec.~\ref{sec:datasets}), amongst other aspects, each speaker is scored for their overall intelligibility on a five-point scale by SLPs. %
In this work, we consider all utterances from all speakers with ratings and develop models to predict the speech intelligibility ratings. %

\textbf{Tasks.} We train the models on two classification tasks based on the intelligibility ratings for each utterance. First is the \textbf{2-class MILD$+$} task where we predict if the speech sample is \textit{typical} or not (i.e., disordered) by grouping \textit{mild, moderate, severe} and \textit{profound} into the \textit{atypical} class. The second is the \textbf{5-class} task of predicting the 5-point SLP ratings.

\vspace{-0.2cm}
\subsection{SpICE: Speech Intelligibility Classifiers on Euphonia}
Our speech intelligibility classification approach is partly inspired by \cite{venugopalan2021comparing}. They use CNNs, representations from an unsupervised model (TRILL~\cite{shor2021universal}), and representations from an ASR-encoder model to train classifiers on a dataset of 15K utterances focusing on a narrow set of 29 short phrases from each speaker. %
We also use an ASR-encoder,
additionally,
we train a CNN with a learnable frontend~\cite{zeghidour2021leaf} and representations from \wv which uses a transformer backbone. Also, our work significantly scales training, using 550K+ diverse utterances to train classifiers, and extensively evaluates generalization of the models across etiologies and datasets.

\textbf{ASR system encoder representations (ASR-enc).}
This model is identical to that in \cite{venugopalan2021comparing}. We use an LSTM encoder that models acoustic inputs in an ASR system based on an RNN transducer (RNN-T)~\cite{graves2013speech} model. The specific architecture is based on He et.~al.~\cite{he2019streaming} trained on long-form speech~\cite{arun_rnnt}. %
As in \cite{venugopalan2021comparing}, we consider the average-pooled (over time) embeddings of the encoder as the  representation of a speech sample, and train linear models using logistic regression, random forest, and linear discriminant analysis on the embeddings to predict class scores.

\textbf{\wv representations.}
To compare ASR-enc with a similarly powerful model,
we train linear classifiers using the self-supervised representations from the final layer of the \wv model~\cite{baevski2020wav2vec} publicly available on HuggingFace~\cite{wolf2020transformers}. This architecture consists of a multi-layer CNN that produces latent speech representations of raw audio, and uses a transformer~\cite{vaswani2017attention} and masked language modeling~\cite{devlin2018bert} to build contextualized representations. %
We develop classifiers on representations from the the 12$^{th}$ (768-d) final layer.

\textbf{Fully learnable convolutional classifier (LEAF + CNN)}
As a baseline, we train a fully learnable convolutional classifier. Unlike the CNN classifier of \cite{venugopalan2021comparing}, which takes as inputs fixed mel-filterbanks, the low-level representations of our model are provided by a LEAF \cite{zeghidour2021leaf} frontend which jointly learns filtering, pooling, compression and normalization from data. This frontend feeds into a 2D CNN, based on \cite{tagliasacchi2019self}, which alternates convolutions along time (($3\times1$) kernel) and frequency (($1\times3$) kernel). This is trained using cross-entropy to predict intelligibility on either 2 or 5 classes. %

\section{Datasets}
\label{sec:datasets}
\begin{figure*}[!t]
  \centering
  \includegraphics[width=0.9\textwidth]{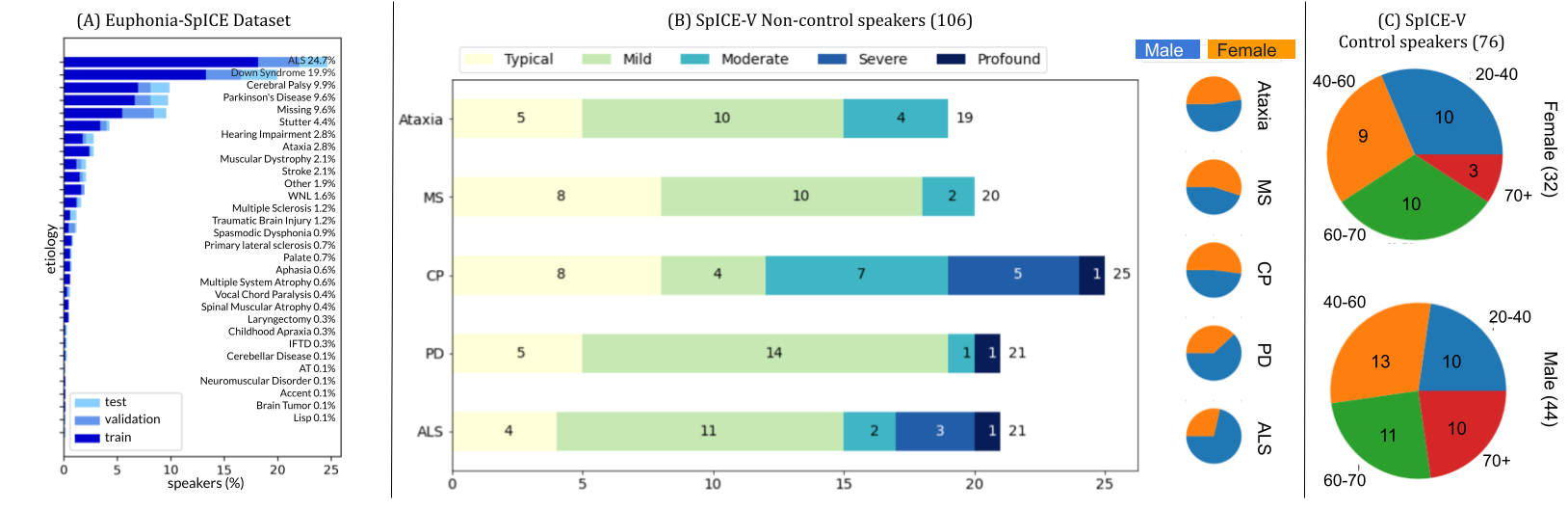}
  \caption{[SpICE datasets] (A) Dist.~of etiologies in the Euphonia-SpICE dataset. (B) SpICE-V non-control speakers split by etiology, intelligibility class and inferred gender, and (C) SpICE-V controls split by inferred age bucket and gender.}
  \label{fig:spiceyt_dist}
  \vspace{-0.2cm}
\end{figure*}

\textbf{Euphonia-SpICE Dataset.} Our training data is a subset of the Euphonia dataset~\cite{euphonia1Mdata}. %
We use data from 677 speakers (756,147 utterances) who were rated by SLPs using a Quality Control (QC) phrase set
of 29 short phrases for each participant. SLPs listened to the QC recordings for each speaker and assessed, among other things, the overall intelligibility of the speaker on a five-point Likert scale. The scale was mapped to 5 classes - \textit{typical, mild, moderate, severe,} and \textit{profound} (detailed in \cite{euphonia1Mdata}). All utterances from a speaker are labeled with the same rating. While \cite{venugopalan2021comparing} only uses the QC utterances, we use the full data ($\approx50\times$) which we call the Euphonia-SpICE dataset. %
The speakers were randomly split into train, val and test set in a 70:15:15 ratio. All our models were trained on the same splits. 
Fig.~\ref{fig:spiceyt_dist}A shows the distribution of etiologies and Tab.~\ref{tab:split-stats} presents the number of speakers and utterances in each split for each label, along with the overall count. %
\begin{table}[t]
\setlength{\tabcolsep}{2pt}
\begin{center}
\caption{Euphonia-SpICE: Count of speakers and utterances}\label{tab:split-stats}
\resizebox{0.8\columnwidth}{!}{
\begin{tabular}{l|ccc|ccc}
\toprule
\multirow{2}{*}{{Intelligibility}} & \multicolumn{3}{c|}{{\# speakers}} & \multicolumn{3}{c}{{\# utterances}} \\
 & \multicolumn{1}{|c}{{Train}} & \multicolumn{1}{c}{{Val.}} & \multicolumn{1}{c}{{Test}} &
\multicolumn{1}{|c}{{Train}} & \multicolumn{1}{c}{{Val.}} & \multicolumn{1}{c}{{Test}} \\
\midrule
TYPICAL &  161 &   41 & 25 & 149,941 & 24,142 & 10,664\\
MILD    &  161 &   29 & 37 & 208,843 & 22,532 & 39,007\\
MODERATE&  83 &   23 & 19 & 124,984 & 48,814 & 21,214\\
SEVERE  &  54 &   12 & 15 & 60,692 & 13,868 & 22,397\\
PROFOUND&  9 &   4 & 4 & 6,716 & 1,691 & 642\\
\midrule
OVERALL &  468 &   109 & 100 &  551,176 & 111,047 & 93,924  \\
\bottomrule
\end{tabular}
}
\end{center}
\vspace{-0.8cm}
\end{table}

\subsection{Datasets for evaluating generalization.}
We evaluate our trained models (inference only) on multiple datasets to demonstrate flexibility and generalizability of the approach to diverse disorders and data collection setups.

\textbf{UASpeech}~\cite{kim2008dysarthric} is a database of dysarthric speech produced by speakers with CP. Our academic collaborator obtained access %
to the data and evaluated our \wv model on ``all words'' (765 utterances per speaker) from 28 consented speakers (15 dysarthric and 13 controls). We used audio from channel 5 of the 8-microphone array of recordings. %

\textbf{TORGO.}  
We use a subset of \torgo~\cite{rudzicz2012torgo}, as described in \cite{8962210}. In particular, the subset of control speech sentences also available as dysarthric speech, and we use only the recordings from the microphone array.
This yielded ~1200 utterances across 7 dysarthric speakers and 7 matched controls with intelligibility labels in [a,b,c,d,e] (`a' being the most intelligible and `e' being least). Additionally, we had one SLP rate each speaker on the same Likert scale as in the Euphonia dataset. The SLP listened to 10 utterances for each speaker (selected randomly with no overlap in the phrases between speakers) and was asked to rate the overall intelligibility of the speaker on the five-point scale.

\textbf{ALS-TDI PMP} dataset~\cite{vieira2022machine} was collected from over 500 people living with ALS over a 4 year period. Participants recorded voice samples\footnote{They repeat the phrase `I owe you a yo-yo today` five times} and self-reported ALS Functional Rating Scale (ALSFRS-R) scores (integer in [0-4]) for 12 functions one of which is speech. We use the test split from \cite{vieira2022machine} consisting of 1333 recordings from 90 participants.

\textbf{SpICE-V} To evaluate our models on unprompted speech in realistic settings from speakers with different disorders, we curated our own dataset from a collection of web videos. SLPs identified ${\sim}20$ speaker videos each for 5 etiologies: ALS, CP, PD, Ataxia, and Multiple Sclerosis (MS), accounting for balance in severity and inferred gender. They also marked time segments when the dysarthric speaker was speaking. We gathered control samples from the AudioSet~\cite{audioset} dataset. We watched videos labeled `Male speech' and `Female speech' and selected speakers to balance for inferred age\footnote{Many were public figures (e.g athletes, politicians) with wikipedia pages.} and gender. In total, we collected 106 dysarthric speaker videos containing 2221 utterances (time segments) and 76 control speaker samples (1$\times$10s segments each). The distribution of the data is presented in Fig.~\ref{fig:spiceyt_dist}.

\vspace{-0.3cm}
\section{Results}
\vspace{-0.1cm}
\textbf{Evaluation metrics}
We train and evaluate our models on the Euphonia-SpICE dataset. %
We report utterance-level performances of the models on evaluation metrics used in ~\cite{venugopalan2021comparing}. Namely, \textbf{Accuracy} (Acc.), \textbf{F1 score} and
\textbf{1-vs-rest AUC} (AUC) where we compute the Area Under the Receiver Operating Characteristic Curve for each class against the rest (akin to multi-label classification) and report the mean. %

\vspace{-0.2cm}
\subsection{Euphonia-SpICE performance.} 
\vspace{-0.2cm}
Tab.~\ref{tab:qnt-eval} presents the results of the models when trained and evaluated on the Euphonia-SpICE dataset. %
The ASR-enc model has the best performance on both tasks. The \wv \ based model closely matches ASR-enc performance on the 2-class task; on the 5-class task it does slightly worse. %
The LEAF + CNN model which is far smaller does comparably worse, and we drop it from further evaluations.
\begin{table}[t]
\begin{center}
\caption[qnt-eval]{
[Euphonia-SpICE] We report the mean 1-vs-rest AUC values, F1 score, and accuracy (Acc.) at the utterance-level. %
Higher is better. \textbf{bold} indicates highest value.
}\label{tab:qnt-eval}
\resizebox{\columnwidth}{!}{
\begin{tabular}{l|cc|ccc|ccc}
\toprule
\multicolumn{1}{c|}{{}} & \multicolumn{2}{c|}{{Size}} & \multicolumn{3}{c|}{{\textbf{2-class MILD$+$}}} &
\multicolumn{3}{c}{{\textbf{5-class}}} 
\\
\multicolumn{1}{c|}{{Models}} & (MB) & params &
AUC & F1 & Acc. & AUC & F1 & Acc.\\
\midrule
LEAF + CNN & 55 & 8M &
0.669 & 0.833  & \textbf{0.886} &  
0.600 & 0.362  & 0.378 \\
\wv & 360 & 100M &
0.742 & 0.857 & 0.863 &
0.652 & 0.416 & 0.423 \\
ASR-enc & 122 & 60M &
\textbf{0.761} & \textbf{0.861} & 0.862 &
\textbf{0.714} & \textbf{0.422} & \textbf{0.432} \\
\bottomrule
\end{tabular}
}
\end{center}
\vspace{-0.7cm}
\end{table}

\vspace{-0.2cm}
\subsection{Models generalize well on existing datasets.}
\textbf{TORGO.} Tab.~\ref{tab:torgo-res} presents results %
on \torgo . We compute  predictions at the speaker-level, by averaging the 5-class scores of the model across all utterances and pick the argmax. We present the utterance-level accuracy (in parenthesis) on the binary classification task of whether the (5-class) model correctly identifies each utterance from a speaker as either \textit{typical} or not as determined by our SLP. We observe that both the ASR-enc model and the \wv \ based model trained on the Euphonia-SpICE dataset generalize well. %
\begin{table}[h]
\setlength{\tabcolsep}{2pt}
\begin{center}
\caption[torgo]{[TORGO] Generalization (only inference) on TORGO. Per-speaker predictions and (binarized accuracy \%).}\label{tab:torgo-res}
\resizebox{\columnwidth}{!}{
\rowcolors{2}{gray!20}{gray!2}
\begin{tabular}{cccc|lll}
\toprule
\rowcolor{white} %
& & TORGO & SLP & \multicolumn{3}{c}{SpICE 5-cls models} \\
Speaker & \# Utts. & label & label & LEAF+CNN & \wv & ASR-enc \\
\midrule
 FC01 &  26 & Control & {\textcolor{myb}{ typical}} & {\textcolor{myb}{typ.}} (34.6) & {\textcolor{myb}{ typ.}} (96.2) & {\textcolor{myb}{typ.}} (96.2) \\
 FC02 & 122 & Control & {\textcolor{myb}{ typical}} & {\textcolor{myb}{typ.}} (68.9) & {\textcolor{myb}{ typ.}} (95.9) & {\textcolor{myb}{typ.}} (100)  \\
 FC03 & 125 & Control & {\textcolor{myb}{ typical}} & {\textcolor{myb}{typ.}} (65.6) & {\textcolor{myb}{ typ.}} (83.2) & {\textcolor{myb}{typ.}} (78.4) \\
 MC01 & 118 & Control & {\textcolor{myb}{ typical}} & {\textcolor{myb}{typ.}} (55.1) & {\textcolor{myb}{ typ.}} (96.6) & {\textcolor{myb}{typ.}} (92.4) \\
 MC02 & 122 & Control & {\textcolor{myb}{ typical}} & {\textcolor{myr}{sev.}} (22.1) & {\textcolor{myb}{ typ.}} (94.3) & {\textcolor{myb}{typ.}} (92.6) \\
 MC03 & 119 & Control & {\textcolor{myb}{ typical}} & {\textcolor{myb}{typ.}} (75.6) & {\textcolor{myb}{ typ.}} (98.3) & {\textcolor{myb}{typ.}} (98.3) \\
 MC04 & 121 & Control & {\textcolor{myb}{ typical}} & {\textcolor{myo}{mod.}} (5)    & {\textcolor{myb}{ typ.}} (98.3) & {\textcolor{myb}{typ.}} (99.2) \\
  F03 & 100 &       a & {\textcolor{myy}{    mild}} & {\textcolor{myb}{typ.}} (63)   & {\textcolor{myy}{ mild}} (87.0) & {\textcolor{myy}{mild}} (88.0) \\
  F04 &  97 &       a & {\textcolor{myb}{ typical}} & {\textcolor{myo}{mod.}} (8.2)  & {\textcolor{myb}{ typ.}} (91.8) & {\textcolor{myb}{typ.}} (74.2) \\
  M03 &  92 &       a & {\textcolor{myb}{ typical}} & {\textcolor{myo}{mod.}} (15.2) & {\textcolor{myb}{ typ.}} (98.9) & {\textcolor{myb}{typ.}} (100)  \\
  F01 &  20 &     d/e & {\textcolor{myo}{moderate}} & {\textcolor{myo}{mod.}} (85)   & {\textcolor{myo}{ mod.}} (100) & {\textcolor{myo}{mod.}} (100)  \\
  M02 &  92 &     d/e & {\textcolor{myo}{moderate}} & {\textcolor{myo}{mod.}} (92.4) & {\textcolor{myy}{ mild}} (100) & {\textcolor{myy}{mild}} (100)  \\
  M04 &  86 &     d/e & {\textcolor{myr}{  severe}} & {\textcolor{myo}{mod.}} (59.3) & {\textcolor{myr}{ sev.}} (100) & {\textcolor{myo}{mod.}} (100)  \\
  M05 &  17 &       c & {\textcolor{myr}{  severe}} & {\textcolor{myb}{typ.}} (41.2) & {\textcolor{myr}{ sev.}} (100) & {\textcolor{myo}{mod.}} (100)  \\
\bottomrule
\end{tabular}
} 
\end{center}
\vspace{-0.5cm}
\end{table}

\textbf{UASpeech}
 contains speaker-level intelligibility ratings in the 1-100\% range. We use a simple map from predicted class to intelligibility \{0:100\%, 1: 90\%, 2: 60\%, 3: 40\%, 4: 20\%\} and average predictions across utterances (the mapping didn't seem to matter as long as it was monotonic).
 In Tab.~\ref{tab:uaspeech_corr} we compare performance with prior work~\cite{tripathi2020novel} which uses an ASR model's error rates that requires transcription, and measures Pearson correlation between predictions and labels. 
 {\footnotesize{\hspace{0.2cm} Due to UASpeech access restrictions, we only evaluate on \wv.}}

\begin{table}[h]
\setlength{\tabcolsep}{2pt}
\begin{center}
\caption[UASpeech]{Pearson correlation on  UASpeech.}\label{tab:uaspeech_corr}
\resizebox{\columnwidth}{!}{
\begin{tabular}{l|ccc}
\toprule
Data subset & \# Speakers & SOTA~\cite{tripathi2020novel} & \wv \\
\midrule
Dysarthric; All words & 15 & 0.94 & 0.91 \\ \hline
Controls + Dysarthric; All words &  28 & - & 0.93 \\
\bottomrule 
\end{tabular}
}
\end{center}
\vspace{-0.7cm}
\end{table}

\textbf{Generalization to Speech ALSFRS-R prediction.}
Tab.~\ref{tab:alsfrs_auc}  presents the AUC at the utterance level on predicting the Speech ALSFRS-R score (also 5 classes). We compare with \cite{vieira2022machine} which uses a CNN similar to \cite{venugopalan2021comparing} trained on 3776 recordings from 389 speakers specifically to predict Speech ALSFRS-R scores, whereas we do not do any further training.

\begin{table}[h]
\setlength{\tabcolsep}{2pt}
\begin{center}
\caption[ALSFRS]{AUC on utterance-level ALSFRS-R prediction.}\label{tab:alsfrs_auc}
\begin{tabular}{cc|ccc}
\toprule
\# Spkr & \# Utt. & SOTA~\cite{vieira2022machine} & ASR-enc & \wv \\ \hline
90 & 1333 & 0.86 & 0.82 & 0.81 \\
\bottomrule
\end{tabular}
\end{center}
\vspace{-1.0cm}
\end{table}

\subsection{\wv generalizes well on SpICE-V}
Tab.~\ref{tab:spiceyt-res} presents speaker and utterance-level accuracy of the models. The performance is split based on controls (all of whom have typical speech),  non-controls, and all. We separate out the group that does not include any Dysarthric speakers labeled `Typical' and one which includes all Dysarthric speakers.
In this more challenging dataset we can see that the self-supervised representations from \wv help the model generalize better than the ASR encoder based model.

\begin{table}[h]
\setlength{\tabcolsep}{2pt}
\begin{center}
\caption{[SpICE-V] Comparing \wv  and ASR-enc on speaker- and utterance-level accuracies.
}\label{tab:spiceyt-res}
\resizebox{\columnwidth}{!}{
\rowcolors{2}{gray!20}{gray!2}
\begin{tabular}{lccc|cc|cc}
\toprule
\rowcolor{white} %
        & w. Typ. &  & Total (Atyp.) & \multicolumn{2}{c|}{\wv~Acc.~(\%)} & \multicolumn{2}{c}{{ASR-enc~Acc.~(\%)}} \\
Group & non-ctrl & \# Utts.    &  \# Spkr    &  spkr & utt.   &  spkr & utt. \\
\midrule
Controls     & $\times$     &  76 & 76 (0)  & 76.32 & 76.32 & \textbf{96.42} & \textbf{96.42}   \\
Dysarthric (-Typ.)  & $\times$  &  1489 & 76 (76)  & \textbf{93.42} & \textbf{94.83} & 63.16 & 66.92   \\
Dysarthric (all) & \checkmark   &  2221 & 106 (76)  & \textbf{77.36} & \textbf{75.64} & 68.65 & 67.92   \\
All (-Typ.\&~Dys.)   & $\times$      &  1565 & 152 (76) & \textbf{84.87} & \textbf{93.93} & 78.29 & 68.21   \\
All & \checkmark &  2297 & 182 (76) & 76.92 & \textbf{75.66} & \textbf{78.57} & 69.47   \\
\bottomrule
\end{tabular}
}
\end{center}
\vspace{-0.5cm}
\end{table}

\vspace{-0.4cm}
\section{Discussion}
\vspace{-0.2cm}
\textbf{Models do well on ALS, PD, CP and Ataxia.} 
ALS and CP are the most prevalent in the evaluated datasets: TORGO, UASpeech, and ALS-TDI PMP; and our models do well on these.
When we look at performance sliced by Etiology on SpICE-V (Tab.~\ref{tab:spiceyt-nonctrl-res}) and on the most prevalent 7 etiologies in Euphonia-SpICE test set (Tab.~\ref{tab:etio-res}), we can see at the speaker level the model does well on ALS, CP, PD and Ataxia. The performance on MS is mixed, and the model has difficulty identifying speakers with MS having typical speech.

\begin{table}[h]
\setlength{\tabcolsep}{2pt}
\begin{center}
\caption[Etiology]{[Euphonia-SpICE] Performance sliced by etiology. Both models show similar per-speaker accuracy. %
}\label{tab:etio-res}
\resizebox{\columnwidth}{!}{
\rowcolors{2}{gray!20}{gray!2}
\begin{tabular}{lcc|ccc|c}
\toprule
\rowcolor{white} %
         &  & Atyp./Total & \multicolumn{3}{c|}{{per-utterance AUC}} & Spkrs. \\
Etiology & \# Utts. (\%)    &  \# Spkr    &  & \wv  &  ASR-enc. & Acc\\
\midrule
ALS          &  22076 (23.7) & 14 / 18  & & 0.749 & 0.763 & 0.778   \\
 CP          &  14518 (15.6) & 11 / 12  & & 0.890 & 0.916 & 0.834   \\
Down Syn.    &  13971 (15.0) & 18 / 23  & & 0.544 & 0.525 & 0.652   \\
PD           &  13863 (14.9) & \ 8 / 11 & & 0.489 & 0.521 & 0.727   \\
Hearing Imp. &  8478  ( 9.1) & \ 5 / \ 5& & NA    & NA    & 1.000   \\
MS           &  6272  ( 6.7) & \ 3 / \ 4& & 0.842 & 0.942 & 0.750   \\
Musc. Dystr. &  2544  ( 2.7) & \ 1 / \ 3& & 0.935 & 0.958 & 0.667   \\
\bottomrule
\end{tabular}
}
\end{center}
\vspace{-0.5cm}
\end{table}

\begin{table}[h]
\setlength{\tabcolsep}{2pt}
\begin{center}
\caption{[SpICE-V] Slicing performance by etiology.}\label{tab:spiceyt-nonctrl-res}
\resizebox{\columnwidth}{!}{
\rowcolors{2}{gray!20}{gray!2}
\begin{tabular}{lcc|cc|cc}
\toprule
\rowcolor{white} %
        & & \# Spkr & \multicolumn{2}{c|}{\wv~Acc.~(\%)} & \multicolumn{2}{c}{{ASR-enc~Acc.~(\%)}} \\
Etiology & \# Utt.  &  Total (Typ.)   &  spkr & utt.   &  spkr & utt. \\
\midrule
 ALS    & 443 & 21 (4) & 90.5 & 87.6 & 76.2 & 76.0 \\
  PD    & 498 & 21 (5) & 85.7 & 84.9 & 61.9 & 73.0 \\
  CP    & 620 & 25 (8) & 72.0 & 69.8 & 72.0 & 74.5 \\
  MS    & 352 & 20 (8) & 55.0 & 57.5 & 60.0 & 48.6 \\
 Ataxia & 308 & 19 (5) & 84.2 & 75.6 & 68.4 & 62.1 \\
\bottomrule
\end{tabular}
}
\end{center}
\vspace{-0.5cm}
\end{table}

\textbf{Dysarthric speakers with typical speech are harder to classify.} 
From Tab.~\ref{tab:spiceyt-res} we can observe that the models have different thresholds when predicting on dysarthric speakers with typical speech intelligibility. While the ASR-enc model identifies both controls and non-controls with typical speech as `Typical' the \wv model tends to identify them more often as `Mild'. However, when looking at Dysarthric speakers alone (Tab.~\ref{tab:spiceyt-nonctrl-res}) we see that \wv performs consistently well at the speaker and utterance levels. This is explained by the significant difference in training data of the backbone models and their size.

\textbf{Limitations and future work.}
The data in \textit{Project Euphonia} consists of prompted English speech from participants self-identifying as having speaking disabilities. It has a male-to-female ratio of 60:37 and does not have information on race. Further there is an imbalance across many sensitive etiologies. Future work should consider typical and atypical speech that is more diverse from different demographics, minority groups,  %
and speech in other languages and dialects. Including a fairness testing dataset would also be a valuable contribution. %
To use the models ``in the wild'' it would also be necessary to include non-speech samples and unprompted speech in noisy background. The model can also be fine-tuned and calibrated on a few samples to study applicability to different etiologies.  %

\vspace{-0.3cm}
\section{Conclusion}
\vspace{-0.2cm}
In this work, we trained speech intelligibility classifiers on a large dataset of over half a million utterances from people having a range of speaking disabilities. %
We examined models with different backbones CNNs, LSTMs and transformers. %
We found our classifier to generalize well on several datasets without any additional training and does particularly well on speakers with ALS, CP, PD and Ataxia. %
%

%
\textbf{Acknowledgements.} 
This study would not have been possible without the contributions and efforts of the hundreds of speakers who consented and provided their speech samples through g.co/euphonia, and members of team Euphonia for their data collection effort and feedback.
\bibliographystyle{IEEE}
\small
\bibliography{mybib}

\end{document}